\documentclass[preprint]{elsarticle}
\usepackage[utf8]{inputenc}

\usepackage{hyperref}

\usepackage{lineno}

\usepackage{amsmath}
\usepackage{booktabs} 

\def\useBW{0}

\usepackage[version=4]{mhchem} 

\usepackage{siunitx}
\usepackage{upgreek}
\let\oldequation\equation
\let\oldendequation\endequation
\renewenvironment{equation}
  {\linenomathNonumbers\oldequation}
  {\oldendequation\endlinenomath}

\begin{document}

\begin{frontmatter}

\title{Performance of the GridPix detector quad}

\author[1]{C. Ligtenberg\corref{correspondingauthor}}
\cortext[correspondingauthor]{Corresponding author. Telephone: +31 20 592 2000}
\ead{cligtenb@nikhef.nl}

\author[2]{Y.~Bilevych}
\author[2]{K.~Desch}
\author[1]{H.~van~der~Graaf}
\author[2]{M.~Gruber}
\author[1]{F.~Hartjes}
\author[1,2]{K.~Heijhoff}
\author[2]{J.~Kaminski}
\author[1]{P.M.~Kluit}
\author[1]{N.~van~der~Kolk}
\author[1]{G.~Raven}
\author[2]{T.~Schiffer}
\author[1]{J.~Timmermans}
\address[1]{\normalsize Nikhef, Science Park 105, 1098 XG Amsterdam, The Netherlands}
\address[2]{\normalsize Physikalisches Institut, University of Bonn, Nussallee 12, 53115 Bonn, Germany}

\date{\today}

\begin{abstract}
    A gaseous pixel readout module with four GridPix chips, called the quad, has been developed as a building block for a large time projection chamber readout plane. The quad module has dimensions \SI{39.6 x 28.38}{mm} and an active surface coverage of 68.9\%. The GridPix chip consists of a Timepix3 chip with integrated amplification grid and have a high efficiency to detect single ionisation electrons, which enable to make a precise track position measurement.
    A quad module was installed in a small time projection chamber and measurements of \SI{2.5}{GeV} electrons were performed at the ELSA accelerator in Bonn, where a silicon telescope was used to provide a reference track. The error on the track position measurement, both in the pixel plane and drift direction, is dominated by diffusion. The quad was designed to have minimum electrical field inhomogeneities and distortions, achieving systematics of better than \SI{13}{\um} in the pixel plane. The resolution of the setup is \SI{41}{\um}, where the total systematic error of the quad detector is \SI{24}{\um}.
\end{abstract}


\end{frontmatter}


\section{Introduction}
In drift chambers charged particles are identified through ionisation in the gas. 
For the readout of a time projection chamber (TPC) the finest granularity is offered by pixel readouts.
In particular, a GridPix is a CMOS pixel readout chip with an integrated amplification grid added by MEMS postprocessing techniques \cite{Colas:2004ks,Campbell:2004ib}. 
As a result, single ionisation electrons can be detected with great precision, allowing an excellent track position measurement and an estimate of the number of clusters for an energy loss (dE/dx) measurement for particle identification.

The original GridPix using the Timepix chip \cite{Kaminski:2017bgj} has recently been succeeded by a GridPix based on the Timepix3 chip \cite{Poikela:2014joi}. This newer chip offers superior timing, faster readout speed and the possibility to apply time walk corrections. The first results of a single chip detector have been analysed and published in \cite{Ligtenberg:2018a}. Electron diffusion in gas was found to be the dominant error on the track position measurement and systematics in the pixel plane remained below \SI{10}{\um}. Using a truncated sum, an energy loss (dE/dx) resolution of \SI{4.1}{\percent} was found for an effective track length of \SI{1}{m}. The single chip detector was operated reliably in a test beam experiment. However, equipping a large detector surface poses an entirely new challenge.

In order to cover a large detector surface, it is practical to subdivide it into a number of standardized modules. Here we present the design of a quad module with four Timepix3 chips. Because the quad module has all services under the active area, it can be tiled to cover arbitrarily large areas. The performance of a TPC, read out by a single quad module was tested at the ELSA test beam facility in Bonn. Possible applications are in TPCs at future electron-positron colliders, other particle physics experiments and medical imaging such as proton therapy \cite{Kaminski:2017bgj, Krieger:2017yax}.

\section{Quad detector design and construction}

\subsection{The Timepix3 based GridPix}
Here the GridPix consists of a Timepix3 chip with an integrated grid. Directly on the surface of the Timepix3 chip, a \SI{4}{\um} thick silicon-rich silicon nitride protection layer is deposited in order to prevent damage of the readout chip from discharges. 
On top of this \SI{50}{\um} high SU8 pillars are attached that support the \SI{1}{\um} thick aluminium grid that has \SI{35}{\um} diameter circular holes aligned to the pixel input pads. The grid and dykes design was reoptimized:
at four sides the grid ends on a solid SU8 dyke for which on two sides three readout columns were given up.
The Timepix3 chip has a low equivalent noise charge ($\approx$70 e$^-$) and allows for a simultaneous measurement of the Time of Arrival (ToA) and the Time over Threshold (ToT) using a TDC (clock frequency \SI{640}{MHz}) per pixel. For the readout, one out of the eight available links per chip is connected to a speedy pixel detector readout (SPIDR) board \cite{Visser:2015bsa} at a speed of 80 Mbps. The hardware allows reading out at twice this speed.

\subsection{The quad module design and assembly}
\begin{figure}
    \centering
    \if\useBW1
        \includegraphics[width=0.6\textwidth]{img_gray/figure1.pdf}
    \else
        \includegraphics[width=0.6\textwidth]{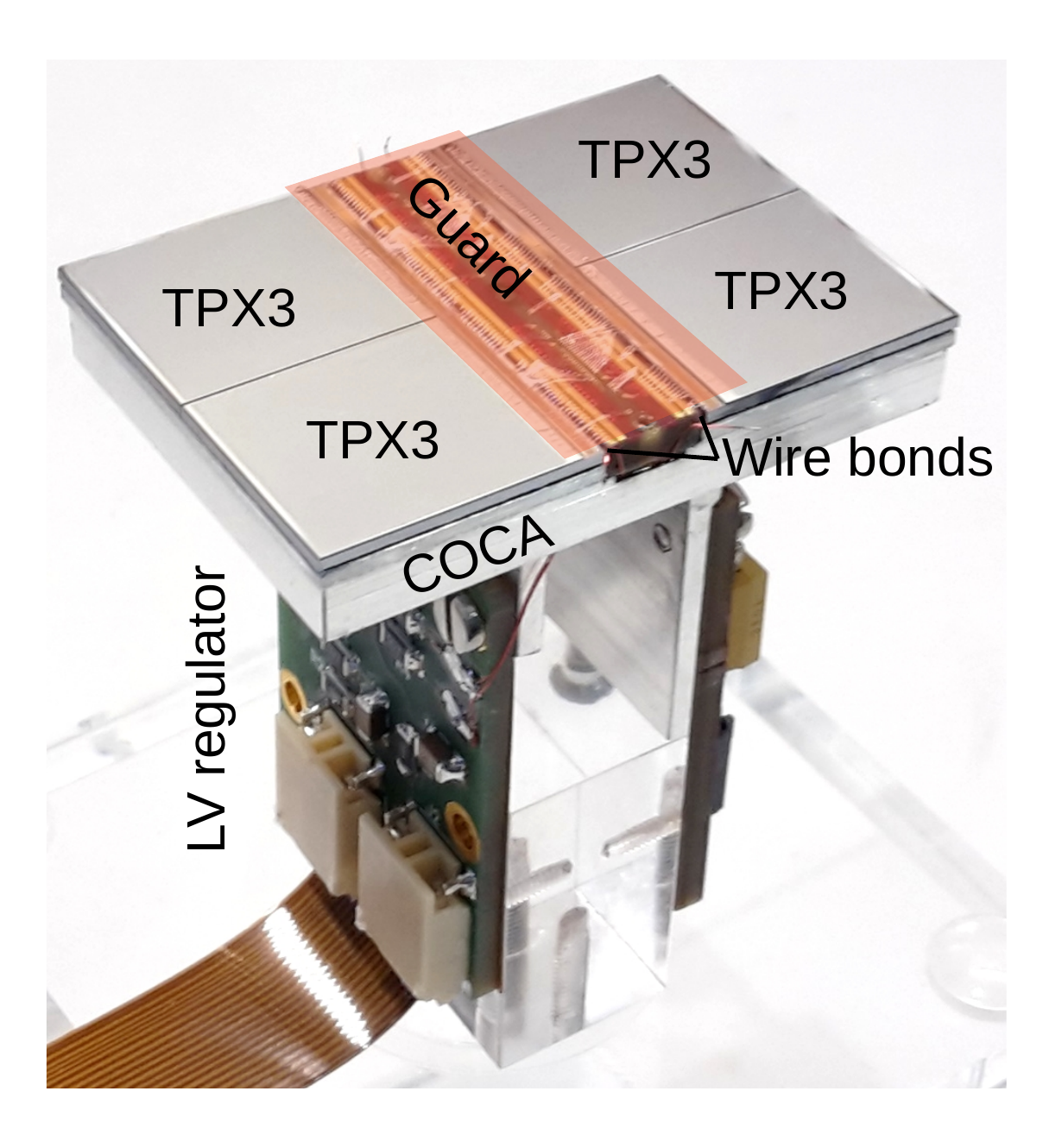}
    \fi
    \caption{Picture of the quad detector with four Timepix3 GridPixes (TPX3) mounted on a cold carrier plate (COCA). The central guard was omitted to show the wire bond PCB, and its operating position is indicated with a transparent rectangle. On the left the Low Voltage (LV) regulator and flexible Kapton cable are visible.}
    \label{fig:quadPicture}
\end{figure}

In order to cover large areas, the quad module shown in figure \ref{fig:quadPicture} was developed. Because of the complexity of the GridPix technology and the fragility of the grids, a small number of four chips per module was chosen. The chips are mounted on a common cooled base plate (COCA). They are electrically connected by wire bonds to a \SI{6}{mm} wide PCB between the two pairs of chips. This allows the control and output lines to be directed to the backside of the quad to maximize the sensitive detection area.  
A short Kapton cable at the other side of the wire bond PCB provides a low impedance connection to the low voltage (LV) regulator. 
The grids are connected by \SI{80}{\um} insulated copper wires to a high voltage (HV) filtering board. The connection to the common HV input uses a 100 M$\Omega$ resistor for each grid to rapidly quench a micro-discharge. 
To support and cool the LV regulator board and the HV filtering board, a U-shaped support is attached by thermally conductive glue under the carrier plate. 
Finally, the wire bonds of the quad are covered by a 10 mm wide central guard electrode located 1.1 mm above the grids to maintain a homogeneous drift field. 

The external quad dimensions are \SI{39.6 x 28.38}{mm} of which 68.9\% is active. In the present design the support components are made of aluminum contributing substantially to the material budget. In the future the material budget can be further minimized by replacing the aluminium by carbon based materials. During low rate operation, the module consumes \SI{8}{W} of power of which \SI{2}{W} in the LV regulator.

\subsection{The quad detector}
The quad module was embedded in a TPC consisting of a steel box and a 40 mm high field cage to provide a homogeneous drift field. The sides of the field cage are formed by \SI{75}{\um} CuBe wires with a \SI{2}{mm} pitch to facilitate UV laser beam measurements. The field cage is terminated on one end by the quad module fitted in a closely surrounding coppered frame at the grid potential, and on the other end by a solid cathode plate. The whole structure was put in a gas-tight container with \SI{50}{\um} Kapton windows on two sides to minimize the material traversed by the beam. 

\section{Test beam measurement}
\begin{figure}
    \centering
    \includegraphics[width=0.8\textwidth]{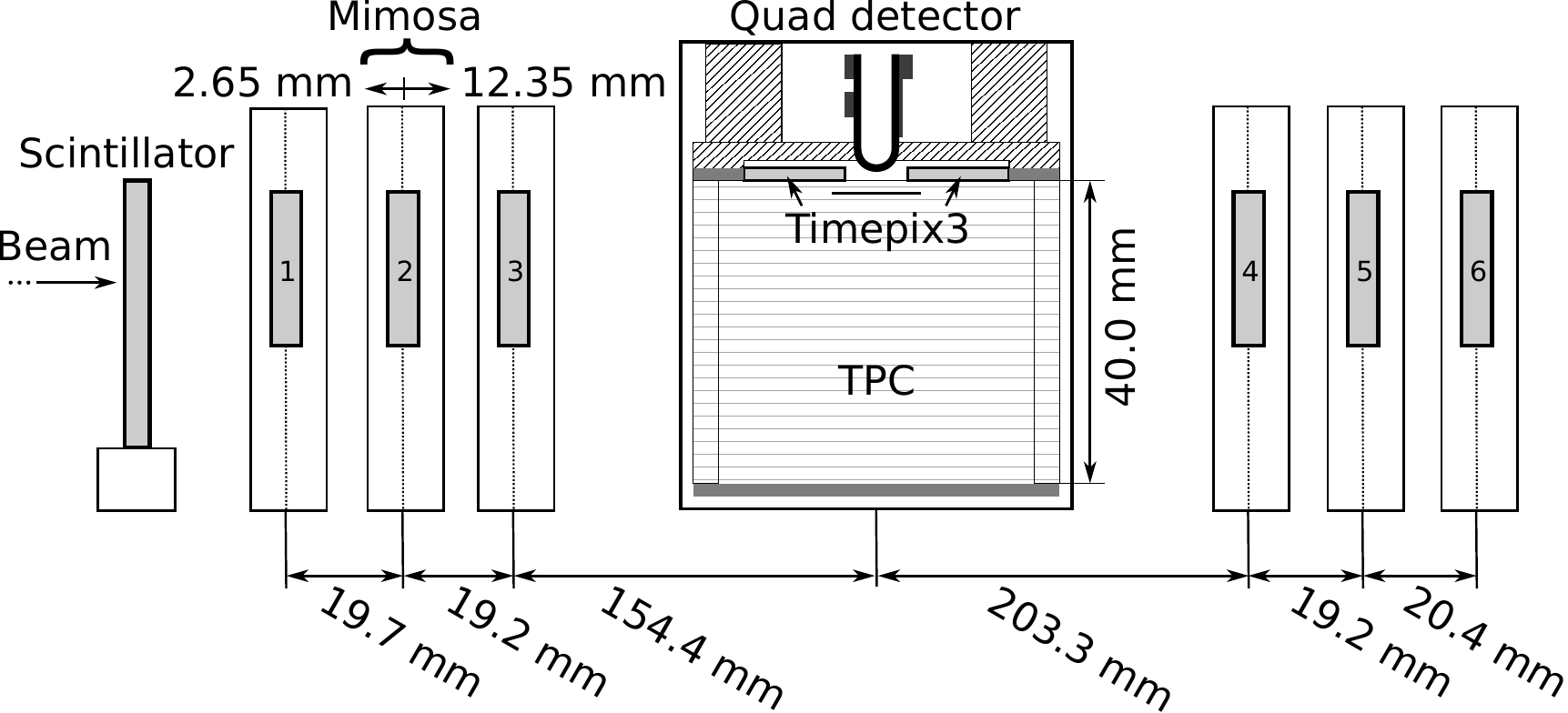} 
    \caption{Setup of the quad with telescope at the ELSA test beam facility.}
    \label{fig:testBeamSetup}
\end{figure}

The device was tested in October 2018 at the ELSA test beam facility in Bonn. The ELSA accelerator provided \SI{2.5}{GeV} electrons at a rate of approximately \SI{10}{kHz} during spills of \SI{16.0}{s} in beam cycles of \SI{17.1}{s}. The whole quad detector was mounted on a remotely controlled slider stage. To provide a precise reference track, the quad TPC was sandwiched between 2 $\times$ 3 planes of a \mbox{Mimosa26 telescope} \cite{Jansen2016}, see figure \ref{fig:testBeamSetup}. Each plane consists of a MAPS detector with $1152\times576$ pixels of size \SI{18.4x18.4}{\um}. 

A scintillator provided a trigger signal to the Trigger Logic Unit (TLU) \cite{TLU} which numbers the triggers and subsequently directs them to both the SPIDR and telescope readout. The telescope hits were collected in time frames of \SI{115.2}{\us}. Due to the high beam intensity, the telescope frames often contain hits from more than one track. The Timepix3 was operated in data driven mode with the trigger data merged in. 

Because of the chosen limited link speed between the Timepix3 chips and the SPIDR a maximum of \SI{1.3}{MHits/s} could be read out. This caused some hits to arrive late at the SPIDR readout, acquiring a wrong \SI{409.6}{\us} course timestamp. As a work-around hits, up to 200 timestamps of \SI{409.6}{\um} after the trigger were collected and analysed. The first track hit had to arrive no more than 5 timestamps late, and the average course timestamp should not deviate more than 150 timestamps. 

During data taking the \SI{700}{ml} gas volume was flushed at a rate of \SI{16.7}{ml/min} with premixed T2K TPC gas. This gas is a mixture consisting of \SI{95}{\percent} \ce{Ar}, \SI{3}{\percent} \ce{CF4}, and \SI{2}{\percent} \ce{iC4H10} suitable for large TPCs because of the relatively high drift velocity and the low diffusion in a magnetic field. The temperature and pressure were relatively stable at \SI{300.5}{K} and \SI{1011}{mbar}. The gas mixture contained a 814 ppm \ce{O2} contamination and a \SI{6000}{ppm} \ce{H2O} contamination, primarily due to the high gas permeability of a silicon rubber cable feed-through.  

The cathode and guard voltages were set such that the electric field was \SI{400}{V/cm}, which is close to the maximum drift velocity for the contaminated gas. The grid voltage was set to \SI{330}{V}, at which there is limited secondary emission from the grid by UV photons produced in avalanches.
The threshold level being a trade off between noise, sensitivity and time walk, was set to about \SI{550}{e^-}.
The gain depends on the beam rate, because of charging up of the protection layer. During the high beam rate, the effective gain was approximately 1000. 
Some of the relevant run parameters are summarized in table \ref{tab:runParameters}.

\begin{table}
    \centering
    \caption{Parameters of the three analyzed runs. The error on the temperature and pressure indicates the spread in the time the three runs were taken.}
    \vspace{3mm}
    \begin{tabular}{l r}
    \toprule
        Runs duration  & 10 minutes \\
        Triggers per run  & \SI{2.2E6}{triggers} \\
        $V_\text{grid}$ & \SI{330}{V}\\
        $E_\text{drift}$ & \SI{400}{V/cm}\\
        Threshold & \SI{550}{e^-}\\
        Temperature & $(300.5\pm0.13)$~K\\
        Pressure & $(1011\pm0.16)$~mbar \\
        Oxygen concentration & \SI{814}{ppm} \\
        Water vapor concentration & \SI{6000}{ppm} \\
        \bottomrule
    \end{tabular}
    \label{tab:runParameters}
\end{table}

\section{Track reconstruction and event selection}

\subsection{Track reconstruction procedure}
Tracks are reconstructed as straight lines. The $y$-axis is defined roughly in the direction of the beam, and the $x$-axis and $z$-axis are in the horizontal and vertical direction respectively. For the telescope, the $y$-coordinate is taken from the plane position, and the $x$-coordinate and $z$-coordinate correspond to the columns and rows. Apart from a small rotation, the $x$ and $y$-coordinates correspond to the columns and rows of the GridPixes. The $z$-coordinate is the drift length calculated from the ToA and the drift velocity. Tracks are fitted using a linear regression fit with hit errors in the two directions perpendicular to the beam $\sigma_x$ and $\sigma_z$. The expressions for the error values are given in sections \ref{sec:hitResolutionx} and \ref{sec:hitResolutionz}.

The detectors are aligned using the data. First, the telescope is independently aligned. The positions in the $y$-direction along the beam are measured and kept fixed. 
Taking one plane as a reference, the other five planes can be rotated. These rotations and additionally two shifts in $x$-direction and $z$-direction for four of the planes are iteratively determined from the fitted tracks. 
Next, the quad detector is aligned. Using iterative alignment each chip has three rotations and two shifts in the $x$ and $z$-directions. Additionally, each chip has one parameter describing the angle in the $xz$-plane between the drift direction and the pixel plane. 

\begin{figure}
    \centering
    \if\useBW1
    \includegraphics[width=0.6\textwidth]{img_gray/figure3.pdf} 
    \else
    \includegraphics[width=0.6\textwidth]{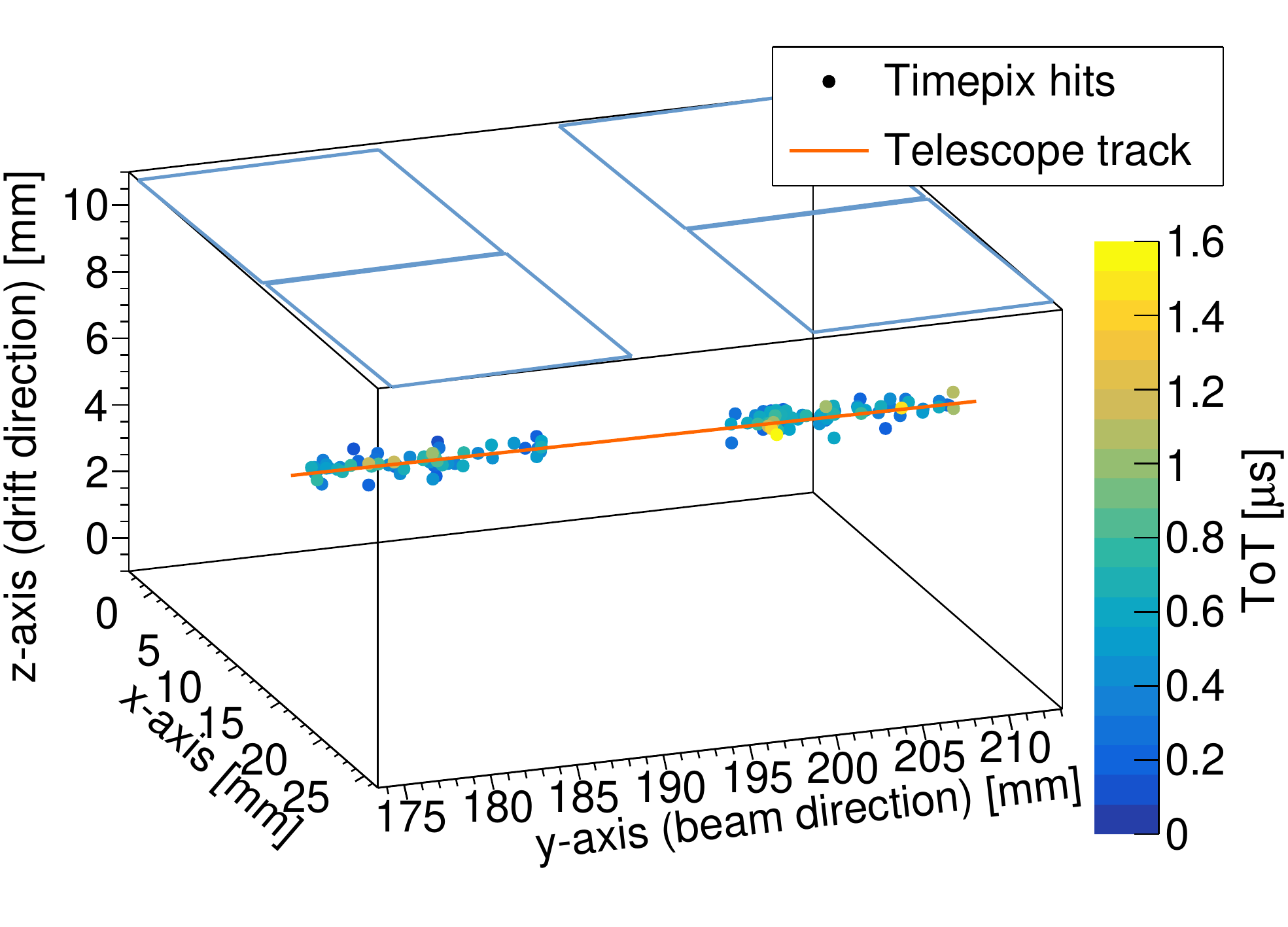} 
    \fi
    \caption{An event with 148 quad detector hits and the corresponding telescope track. The positions of the chips are outlined in blue.}
    \label{fig:eventDisplay}
\end{figure}

An example event with a telescope track is shown in figure \ref{fig:eventDisplay}.

\subsection{Selections}
In the telescope a stringent selection is made to acquire a sample of clean tracks. At least 5 planes should have a hit and the hits should be within \SI{50}{\um} from the track. By requiring the slope difference of the track in the first three planes and in the last three planes to be smaller than \SI{1}{mrad}, scattered tracks are rejected.

GridPix hits are considered if their ToA is within \SI{500}{ns} of a trigger and their ToT is at least \SI{0.15}{\us}. The hits are collected using a track detected by the telescope as a seed. Outliers are rejected by requiring the residuals $r$ (pulls $r/\sigma$) with respect to the telescope track to be smaller than \SI{1.5}{mm} (\SI{2.0}{}) in the $x$-direction and \SI{2.0}{mm} (\SI{3.0}{}) in the $z$-direction.

A track is rejected if it has less than 20 hits. Moreover the average position of all Gridpix hits must be within \SI{0.3}{mm} in the $x$-direction and $z$-direction of the telescope track. Given the high beam rate, the TPC often contains multiple tracks overlapping in time. To suppress overlapping tracks and to reject tracks with delta electrons, \SI{80}{\percent} of the hits within \SI{5}{mm} of a track are required to lie within a distance of \SI{1.5}{mm}. 

The selections are summarized in table \ref{tab:selections} and the total efficiency for events is about 12\%. Most events are rejected, because there are less than 20 GridPix hits corresponding to the telescope track.

\begin{table} 
    \centering
    \caption{Table with selection cuts}
    \vspace{3mm}
    \begin{tabular}{c} \toprule
        Telescope \\ \midrule
        Number of planes hits $\ge$ 5 \\
        Reject outliers ($r_{x,z} <$ \SI{50}{\um}) \\
        Slope difference between sets of planes $<$ \SI{1}{mrad} \\ \midrule
        GridPix hit selection \\ \midrule
        $\SI{-500}{ns} < t_\text{hit}-t_\text{trigger} < \SI{500}{ns} $\\
        Hit ToT $>$ \SI{0.15}{\us} \\
        Reject outliers ( $r_x<\SI{1.5}{mm}, r_z<\SI{2}{mm}$ ) \\
        Reject outliers ( $r_x<2\sigma_x, r_z<3\sigma_z$ ) \\ \midrule
        Event Selection \\ \midrule
        $ N_\text{hits}\ge20 $ \\
        $ ( N_{r_x < 1.5 \text{mm}}$  $/$ $N_{r_x < 5 \text{mm} } )$ $>0.8$ \\
        $|x_\text{Timepix}-x_\text{telescope}|<\SI{0.3}{mm}$ \\
        $|z_\text{Timepix}-z_\text{telescope}|<\SI{0.3}{mm}$ \\
        \bottomrule
    \end{tabular}
    \label{tab:selections}
\end{table}

\section{Results}

\subsection{Number of hits}
\begin{figure}
    \centering
    \if\useBW1
    \includegraphics[width=0.6\textwidth]{img_gray/figure4.pdf}
    \else
    \includegraphics[width=0.6\textwidth]{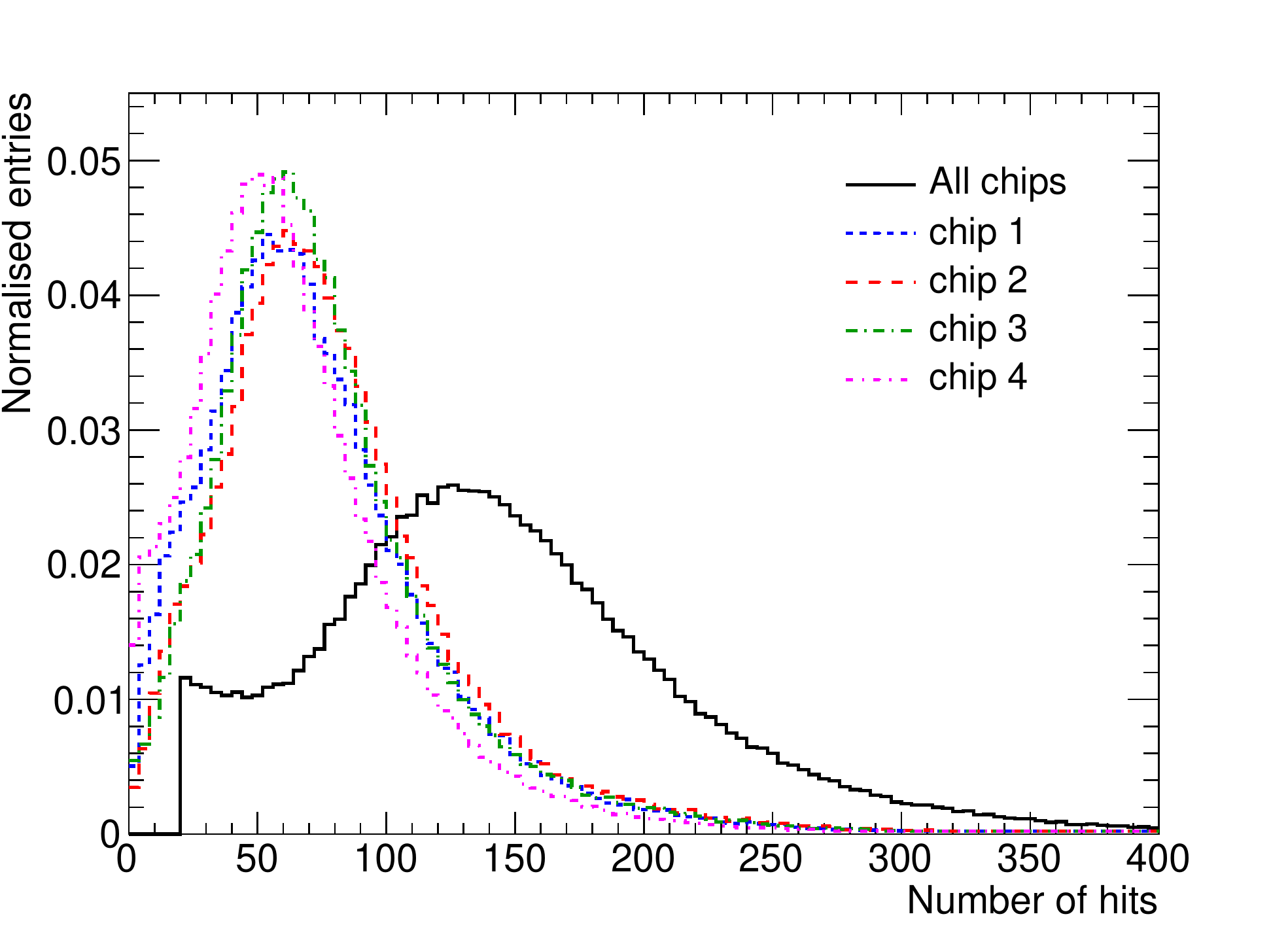}
    \fi
    \caption{Distribution of the number of hits after track selection in total and per chip.}
    \label{fig:nHits}
\end{figure}

The distribution of the number of track hits per chip and the total number of track hits are shown in figure \ref{fig:nHits}. The most probable number of hits per chip varies between 52 and 65 hits, and the mean varies between 65 and 80 hits. The most probable number of hits per quad is 131 and the mean number of track hits is 146 for an effective track length of approximately \SI{27.5}{mm}. This is significantly below the calculated most probable value of 225 electron-ion pairs for a 2.5 GeV electron with this track length  \cite{Garfield}. This is due to the too low effective grid voltage and possibly due to readout problems. Because of the low single electron efficiency, no energy loss (dE/dx) results were extracted.


\subsection{Hit time corrections}
To increase precision in the drift direction, the hit times were corrected. To correct for the double column structure and power distribution deformations of the Timepix3 chip, a ToT factor per column was extracted by injecting test pulses for each pixel. Furthermore, a ToA correction offset was determined from the test beam data based on the underlying substructure of \SI{16x2}{} pixels due to the clock distribution. In addition one ToA correction offset per column and one offset per row was applied. The ToT corrections are of $\mathcal{O}(10\%)$ and the ToA corrections are of $\mathcal{O}(\SI{1}{ns})$.

\subsection{Time walk correction}
\begin{figure}
    \centering
    \if\useBW1
    \includegraphics[width=0.6\textwidth]{img_gray/figure5.pdf}
    \else
    \includegraphics[width=0.6\textwidth]{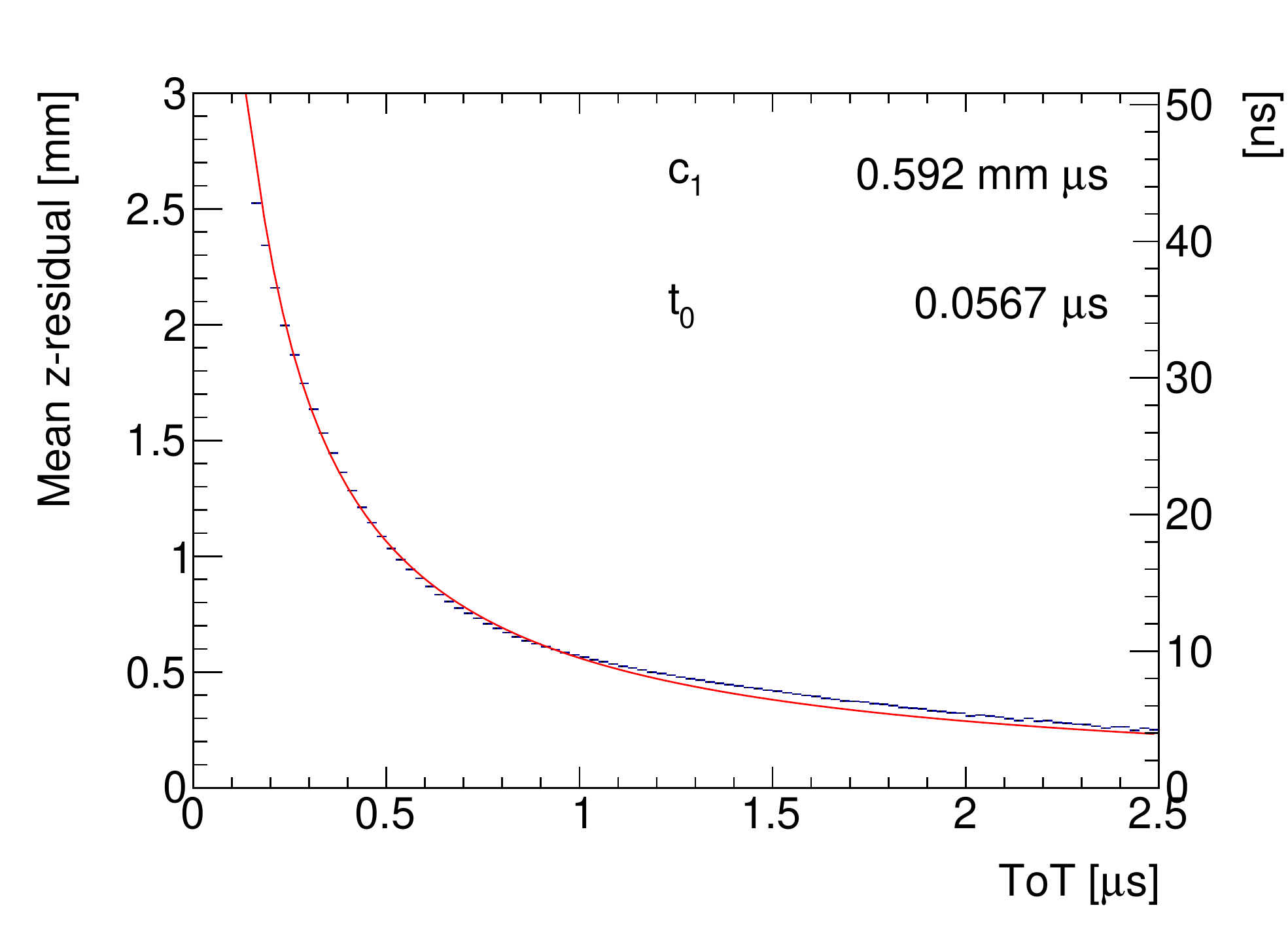}
    \fi
    \caption{Mean z-residual without time walk correction as function of ToT, fitted with equation \eqref{eq:timewalk}. The right axis is given in units of ns using a drift velocity of \SI{59}{\um/ns}.}
    \label{fig:fittedTW}
\end{figure}

A hit is registered when the collected charge reaches the threshold. Since it takes longer for a small signal to reach the threshold than it does for a large signal, the measured ToA depends on the magnitude of the signal. This effect is called time walk and can be corrected by using the ToT as a measure of signal magnitude. In figure \ref{fig:fittedTW} the mean of $z$-residuals is shown as a function of the ToT for all four chips. The relation can be parametrized using the time walk $\delta z_\text{tw}$ as a function of the ToT $t_\text{ToT}$:
\begin{equation}
    \delta z_\text{tw} = \frac{c_1}{t_\text{ToT} + t_0},
    \label{eq:timewalk}
\end{equation}
where $c_1$ and $t_0$ are constants determined from a fit per chip. 

\subsection{Hit resolution in the pixel plane}
\label{sec:hitResolutionx}

\begin{figure}
    \centering
    \if\useBW1
    \includegraphics[width=0.6\textwidth]{img_gray/figure6.pdf}
    \else
    \includegraphics[width=0.6\textwidth]{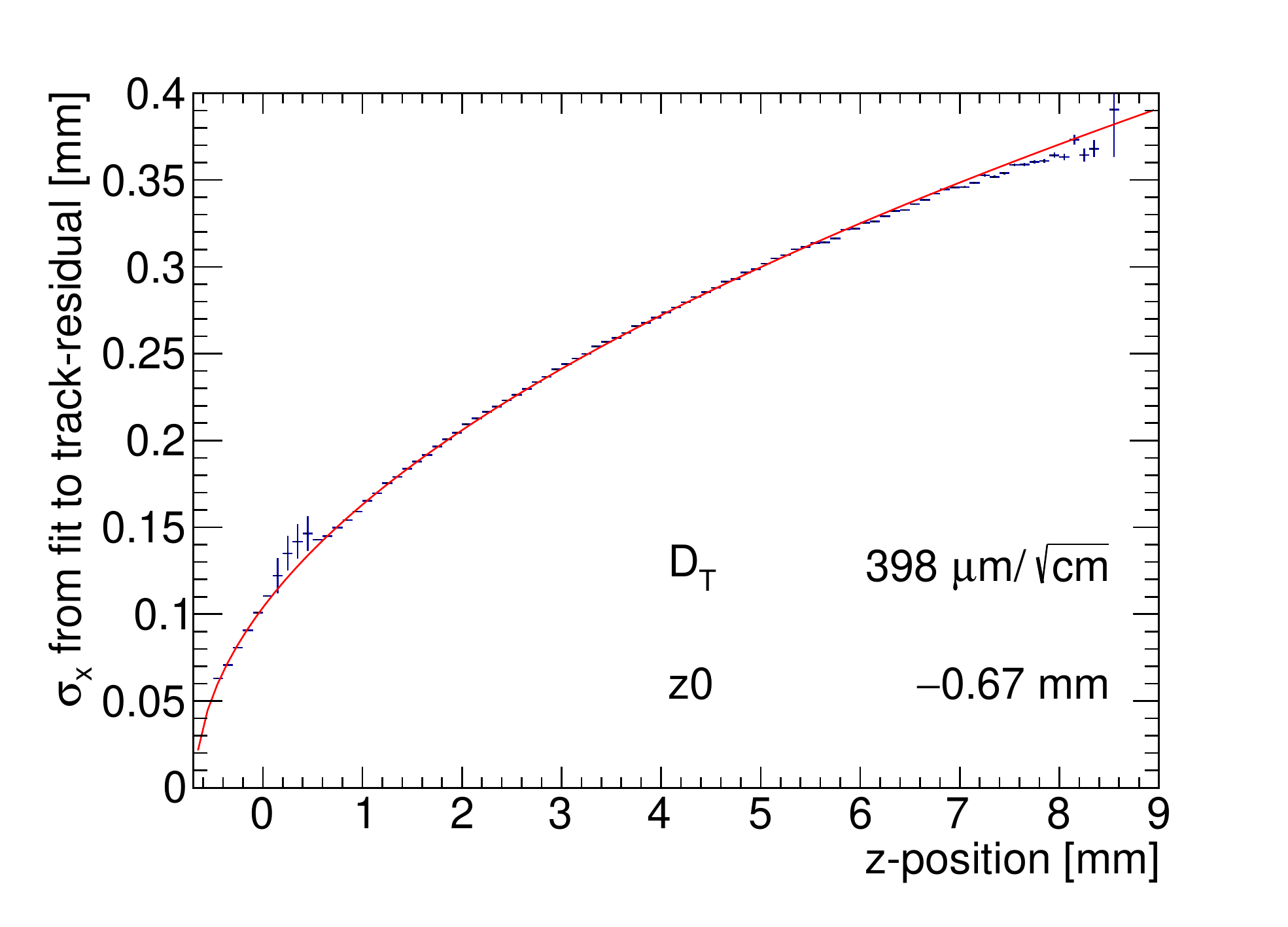}
    \fi
    \caption{Measured hit resolution in the pixel plane (blue points) fitted with the resolution function according to equation \eqref{eq:sigmax} (red line).}
    \label{fig:resolutionx}
\end{figure}

The resolution of the single electrons in the transverse plane ($xy$) was measured as a function of the predicted drift position ($z$). Figure \ref{fig:resolutionx} displays this relation for tracks crossing a fiducial region in the center of the chip. The resolution for the detection of ionisation electrons $\sigma_x$ is  given by:
\begin{equation}
    \sigma_x^2=\frac{d_\text{pixel}^2}{12} +D_T^2(z-z_0),
    \label{eq:sigmax}
\end{equation}
where $d_\text{pixel}$ is the pixel pitch size, $z_0$ is the position of the grid, and $D_T$ is the transverse diffusion coefficient. The resolution at zero drift distance $d_\text{pixel}/\sqrt{12}$ was fixed to \SI{15.9}{\um}.
Tracks with a $z$-position around \SI{0.3}{mm} are given a larger error because they scatter on the central guard. Fitting expression \eqref{eq:sigmax} to the data gives a transverse diffusion coefficient $D_T$ of \SI{398}{\um/\sqrt{cm}} with negligible statistical uncertainty. The measured value is larger than the value of \SI{270}{\um/\sqrt{cm}} $\pm$ 3\% predicted by the gas simulation software Magboltz \cite{Biagi:1999nwa}. Probably this is due to an inaccuracy in the gas mixing, which caused the \ce{CF4} content to be lower than intended.

To compare the precision of the GridPix readout with the precision of conventional pad based TPC readouts, the resolution can be calculated over the length of one pad row. 
For example, at a drift distance of \SI{4}{mm} the resolution of a single ionisation electron is approximately \SI{250}{\um}, so the resolution of a \SI{6}{mm} track segment which has on average 32 electrons is therefore about \SI{44}{\um}.

\subsection{Hit resolution in the drift direction}
\label{sec:hitResolutionz}

The measured $z$-position is directly related to the drift velocity. Using the predicted positions from the telescope, the drift velocity is measured to be \SI{54.6}{\um/ns}, which is slightly lower than the value of \SI{59.0}{\um/ns} expected by Magboltz. Both values have negligible statistical uncertainties.

\begin{figure}
    \centering
    \if\useBW1
    \includegraphics[width=0.6\textwidth]{img_gray/figure7.pdf}
    \else
    \includegraphics[width=0.6\textwidth]{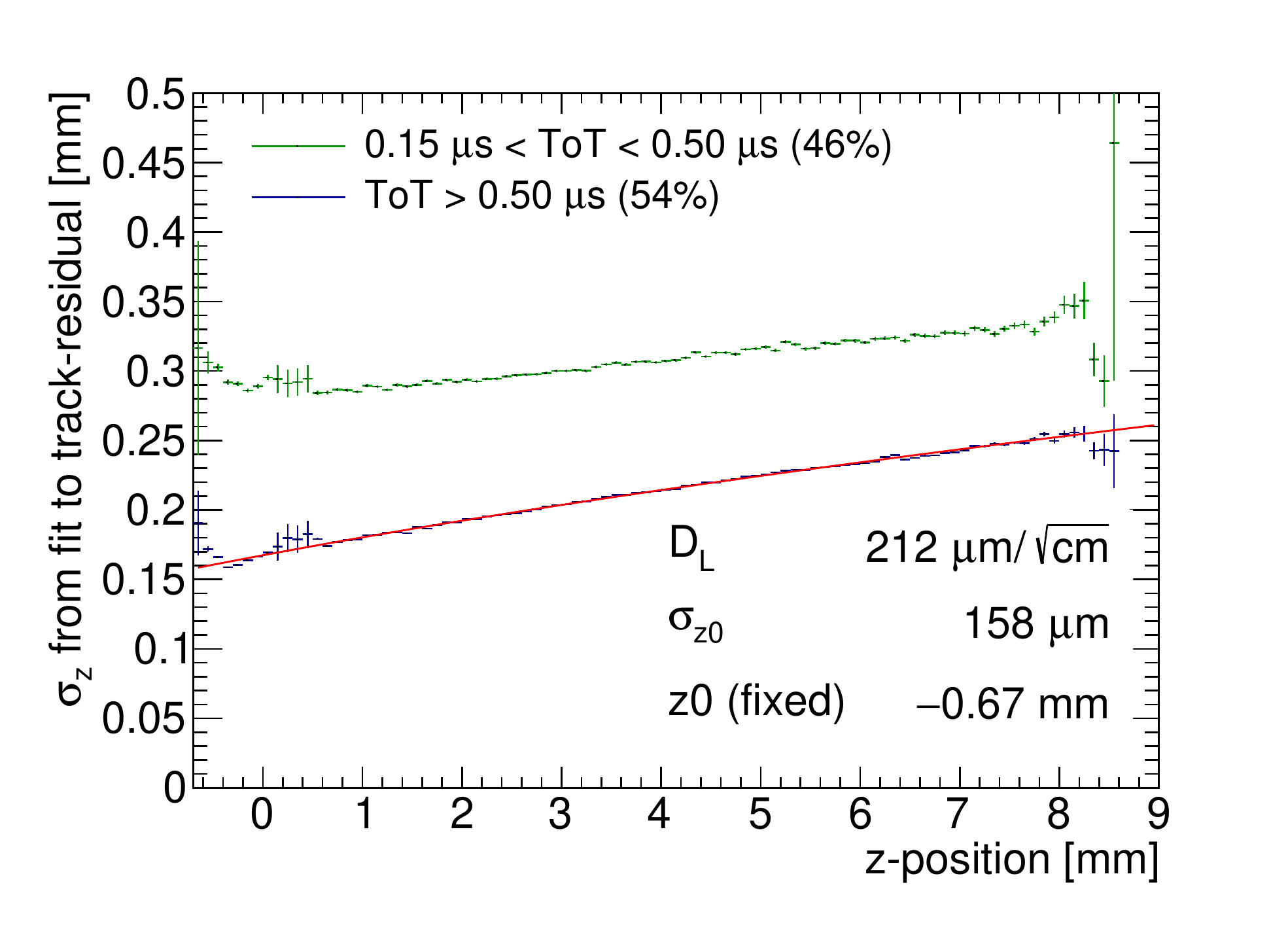}
    \fi
    \caption{Measured hit resolution in drift direction split by ToT. The hits with a ToT above \SI{0.50}{\us} (blue points) are fitted with the resolution according to equation \eqref{eq:sigmaz} (red line). In the legend the fractions of hits in both selections are given.}
    \label{fig:resolutionz}
\end{figure}

The resolution for the detection of ionisation electrons $\sigma_z$ is given by:
\begin{equation}
    \sigma_z^2=\sigma_{z0}^2+D_L^2(z-z_0),
    \label{eq:sigmaz}
\end{equation}
where $\sigma_{z0}$ is the resolution at zero drift distance. The resolution as function of the drift distance is shown in figure \ref{fig:resolutionz} for tracks crossing the fiducial region. Since tracks with a $z$-position around \SI{0.3}{\mm} scatter on the central guard, these data points are given a larger error. The longitudinal diffusion coefficient $D_L$ was determined to be \SI{212}{\um/\sqrt{cm}} with negligible statistical uncertainty, which is equal to the expected value \SI{212}{\um/\sqrt{cm}} $\pm$ 3\% from a Magboltz calculation.

\subsection{Deformations in the pixel plane}
\label{sec:deformationsPixelPlane}
\begin{figure}
    \centering
    \if\useBW1
    \includegraphics[width=0.6\textwidth]{img_gray/figure8.pdf}
    \else
    \includegraphics[width=0.6\textwidth]{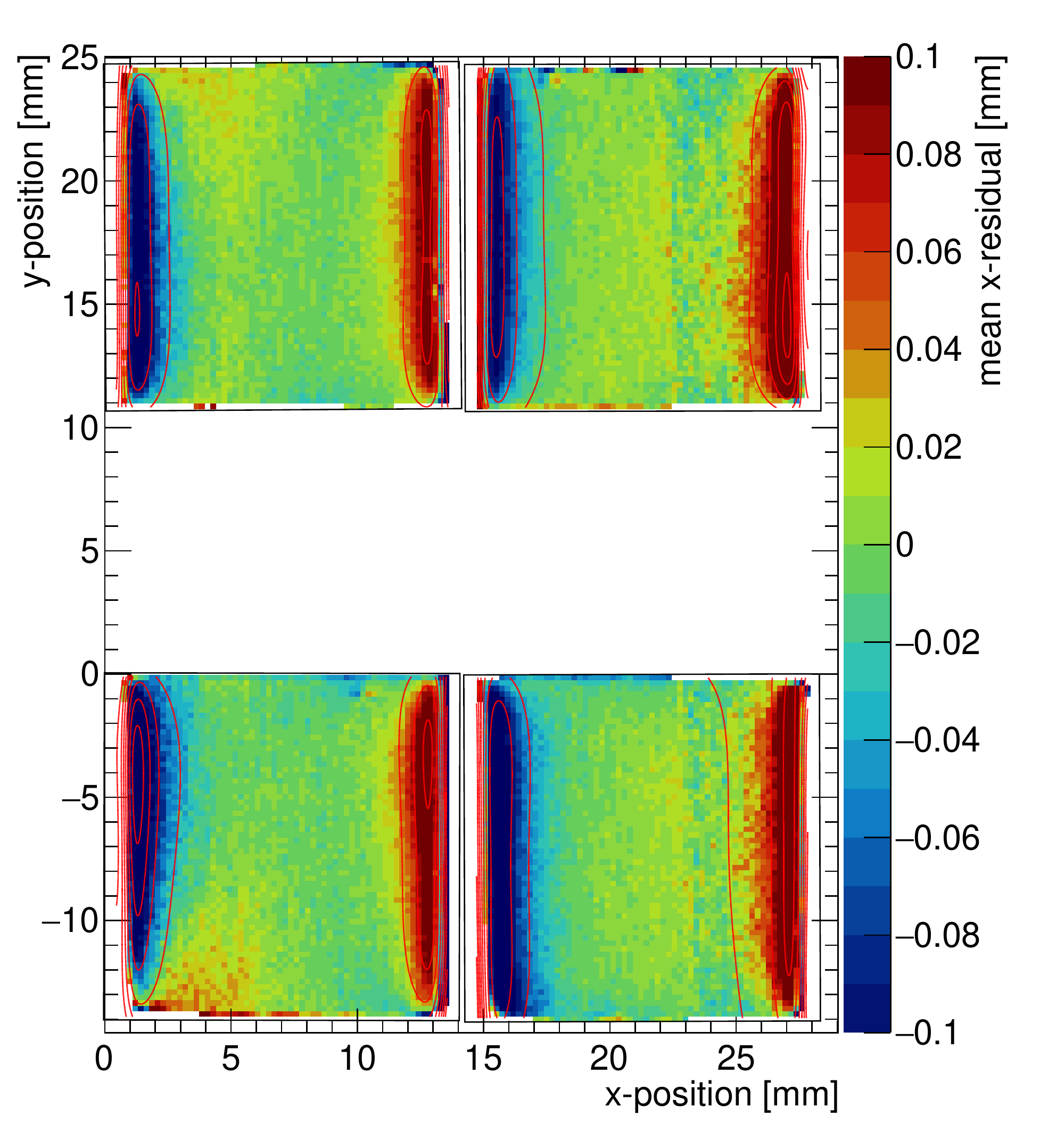}
    \fi
    \caption{Mean residuals in the pixel plane ($x$-residuals) at the expected hit position, fitted with equation \eqref{eq:correction} (red contours).}
    \label{fig:deformationsx}
\end{figure}

It is important to measure possible deformations in the pixel plane ($xy$), because for applications in a TPC this affects the momentum resolution. Because of limited statistics, the mean transverse ($x$) residuals are calculated in bins of \SI{4x4}{} pixels over the quad plane using the tracks defined by the telescope, see figure \ref{fig:deformationsx}. Only bins with more than 800 entries are shown.  

A distortion is present near the edges of the chips. 
The cause is twofold; firstly there is a geometrical bias at the edge of the detector because only part of the ionisation cloud can be detected.  Secondly, the grounded region at the edge of the Timepix3 die causes a non-uniformity of the electric field. 

An empirically selected function of four Cauchy (Breit-Wigner) functions can be fitted to the geometrical bias and the non-uniformity of the field. Near the top and bottom edges the size of the deformations is different, as such - while keeping the other parameters fixed - a $4^\text{th}$ order polynomial function in $y$ was fitted in a second step to set the scale. All in all, the fitted function is given by:
\begin{equation}
    \delta x_\text{deformations} = \sum_{j=0}^4 \left(\frac{1}{\pi}\frac{\gamma_j}{(x-d_{j})^2+\gamma_j^2} \sum_{i=0}^{4} \left( c_{ij} y^i \right)  \right),
    \label{eq:correction}
\end{equation}
where $d_j$ and $\gamma_j$ are the location and scale parameters of the Cauchy distributions. $c_{ij}$ are the parameters of the fourth order polynomial. 

\begin{figure}
    \centering
    \if\useBW1
    \includegraphics[width=0.6\textwidth]{img_gray/figure9.pdf}
    \else
    \includegraphics[width=0.6\textwidth]{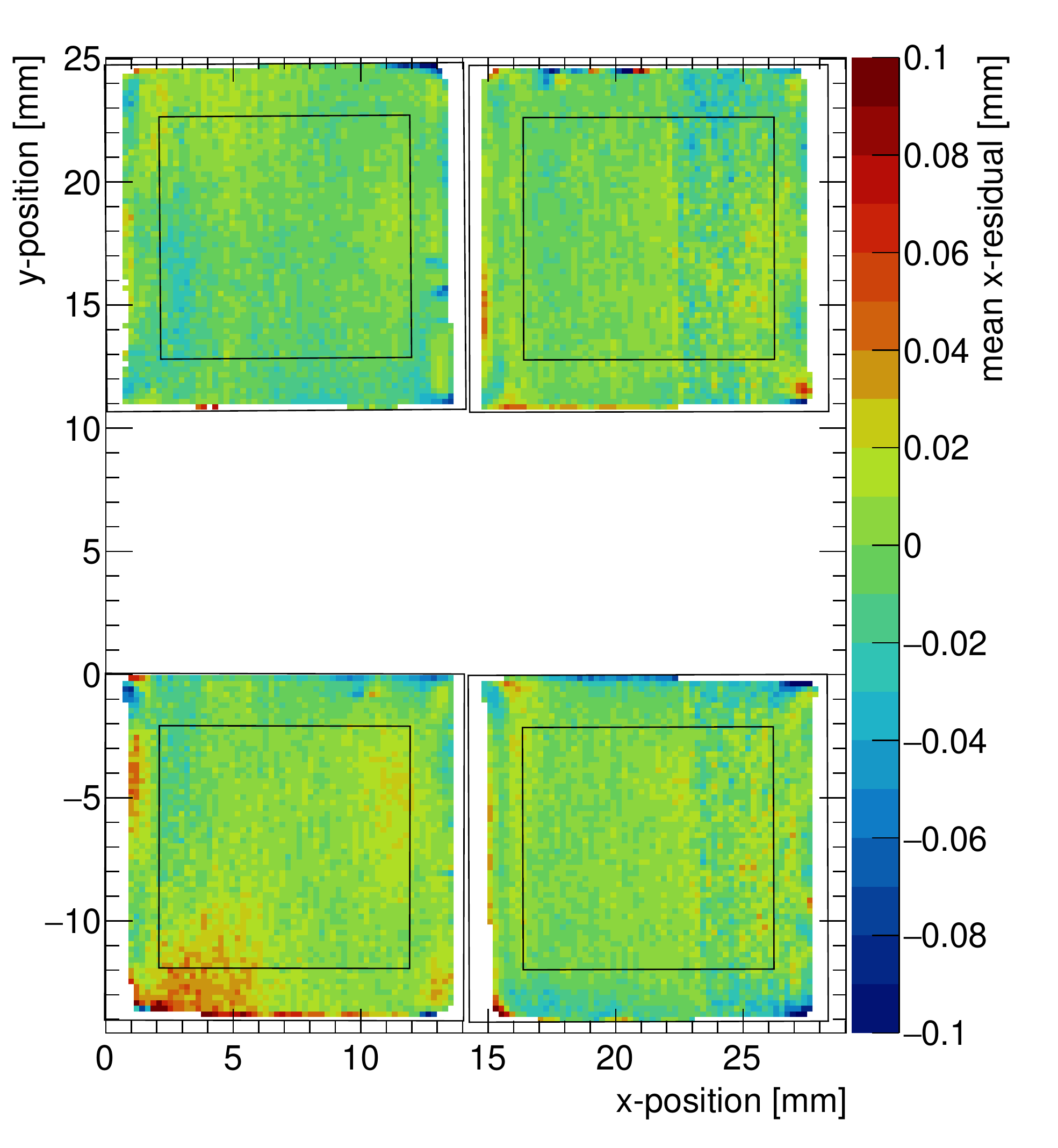}
    \fi
    \caption{mean residuals in the pixel plane ($x$-residuals) at the expected hit position after subtracting the fitted edge deformations using equation \eqref{eq:correction}. }
    \label{fig:deformationsxcorrected}
\end{figure}

The outlines of the fitted function are shown in figure \ref{fig:deformationsx}. The fitted function can be used as a correction by subtracting it from the mean residuals. The result of this procedure is shown in figure \ref{fig:deformationsxcorrected}. 

The r.m.s. of the distribution of the measured mean residual over the surface - or the systematic error for a measurement before the correction in the quad plane - is \SI{31}{\um}. After subtraction of the fitted correction function \eqref{eq:correction}, the r.m.s. of the mean values is \SI{13}{\um} over the whole plane and \SI{9}{\um} in the selected region \SI{2}{mm} from the edges indicated by a black outline. The distribution of the mean $x$-residuals after correction are shown in figure \ref{fig:deformationsFreq}. The distortions could be further reduced by improving the homogeneity of the electric field near the dyke e.g. by adding a field wire above the quad detector at the boundaries between the neighbouring chips.

\subsection{Deformations in the drift direction}
\label{sec:deformationsDrift}
\begin{figure}
    \centering
    \if\useBW1
    \includegraphics[width=0.6\textwidth]{img_gray/figure10.pdf}
    \else
    \includegraphics[width=0.6\textwidth]{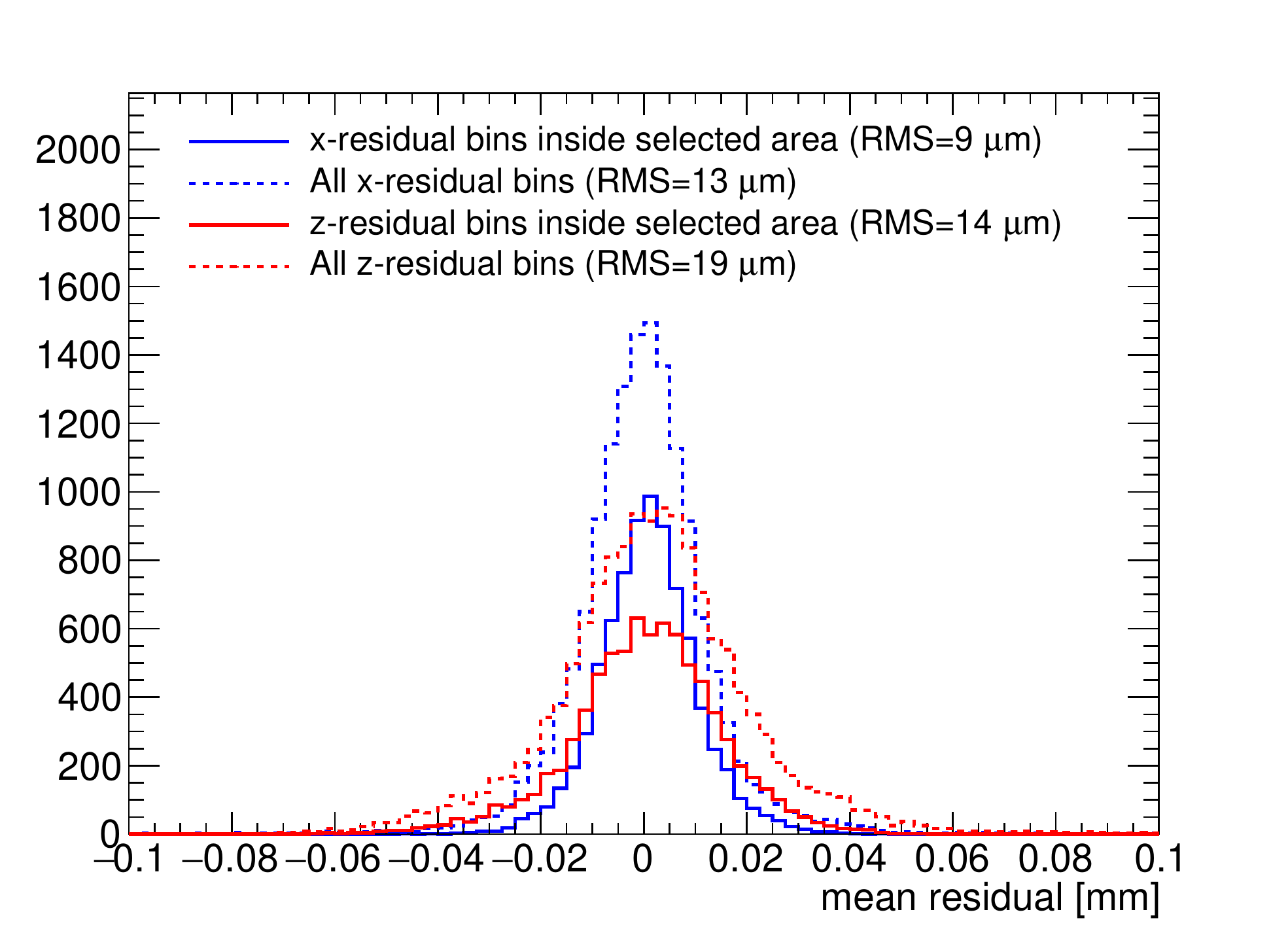}
    \fi
    \caption{Distribution of the mean residuals from $4\times4$ bins in figure \ref{fig:deformationsxcorrected} ($x$-residuals) and figure \ref{fig:deformationsz} ($z$-residuals)}.
    \label{fig:deformationsFreq}
\end{figure}

\begin{figure}
    \centering
    \if\useBW1
    \includegraphics[width=0.6\textwidth]{img_gray/figure11.pdf}
    \else
    \includegraphics[width=0.6\textwidth]{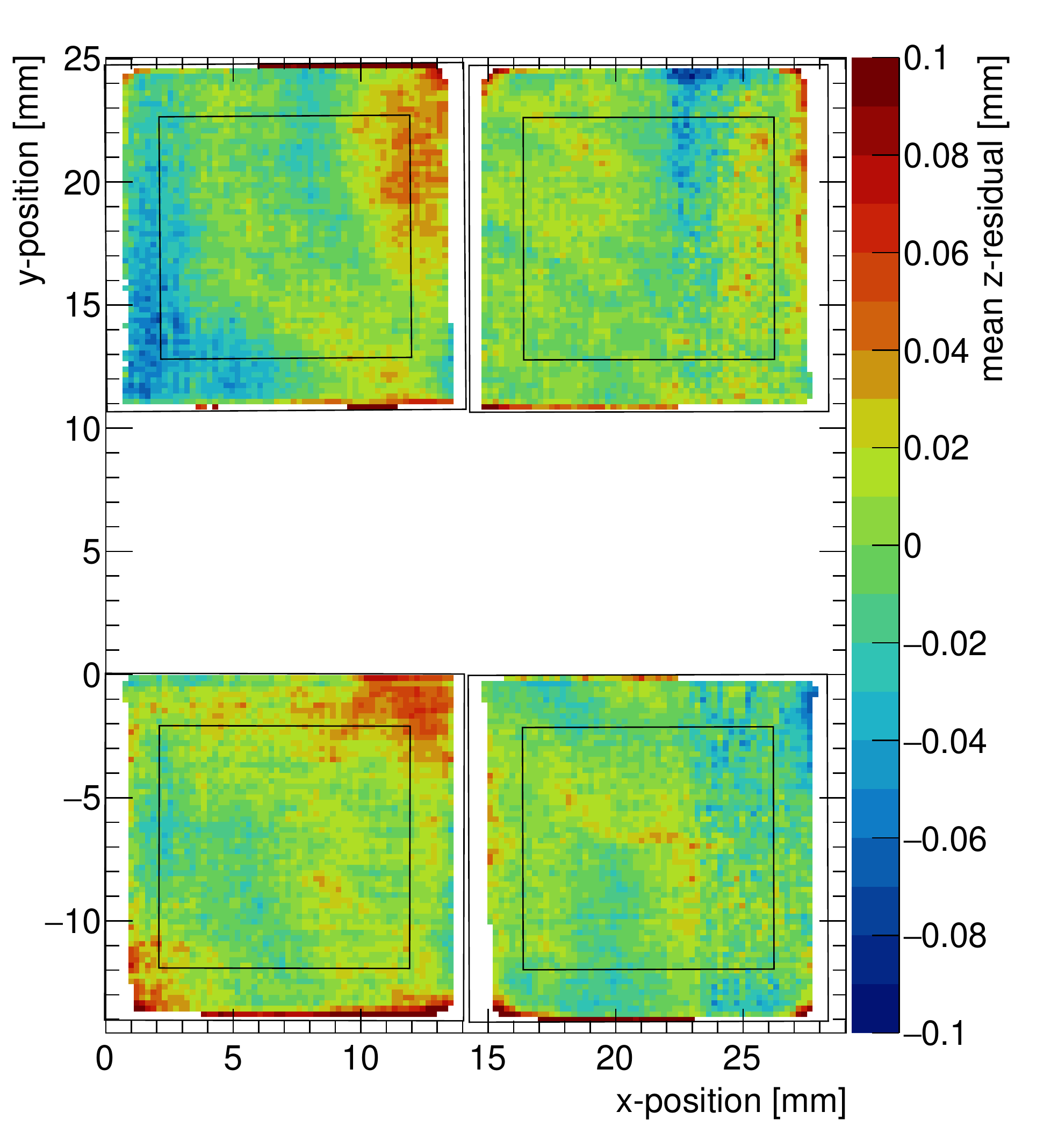}
    \fi
    \caption{Mean residuals in the drift direction ($z$-residuals) at the expected hit position.}
    \label{fig:deformationsz}
\end{figure}

A similar measurement is done for distortions in the drift direction. In figure \ref{fig:deformationsz} the mean longitudinal ($z$) residuals are shown in bins of \SI{4x4}{} pixels over the quad plane using the tracks defined by the telescope. Only bins with more than 800 entries are shown. As shown in figure \ref{fig:deformationsFreq}, the r.m.s. of the distortion is \SI{19}{\um} (\SI{0.35}{ns}) and \SI{14}{\um} (\SI{0.26}{ns}) in the black outlined central area \SI{2}{mm} from the edges.

\subsection{Quad detector resolution}
The overall accuracy of a track position measurement using the quad detector can be tested by comparing the quad track to the telescope track. The difference will be a combination of statistical errors, systematic errors and multiple scattering contributions. Here it is important to estimate the systematical error, because multiple scattering occurs primarily outside the fiducial gas volume, and in applications with multiple quad modules the statistical errors will be further reduced. 

\begin{figure}
    \centering
    \if\useBW1
    \includegraphics[width=0.6\textwidth]{img_gray/figure12.pdf}
    \else
    \includegraphics[width=0.6\textwidth]{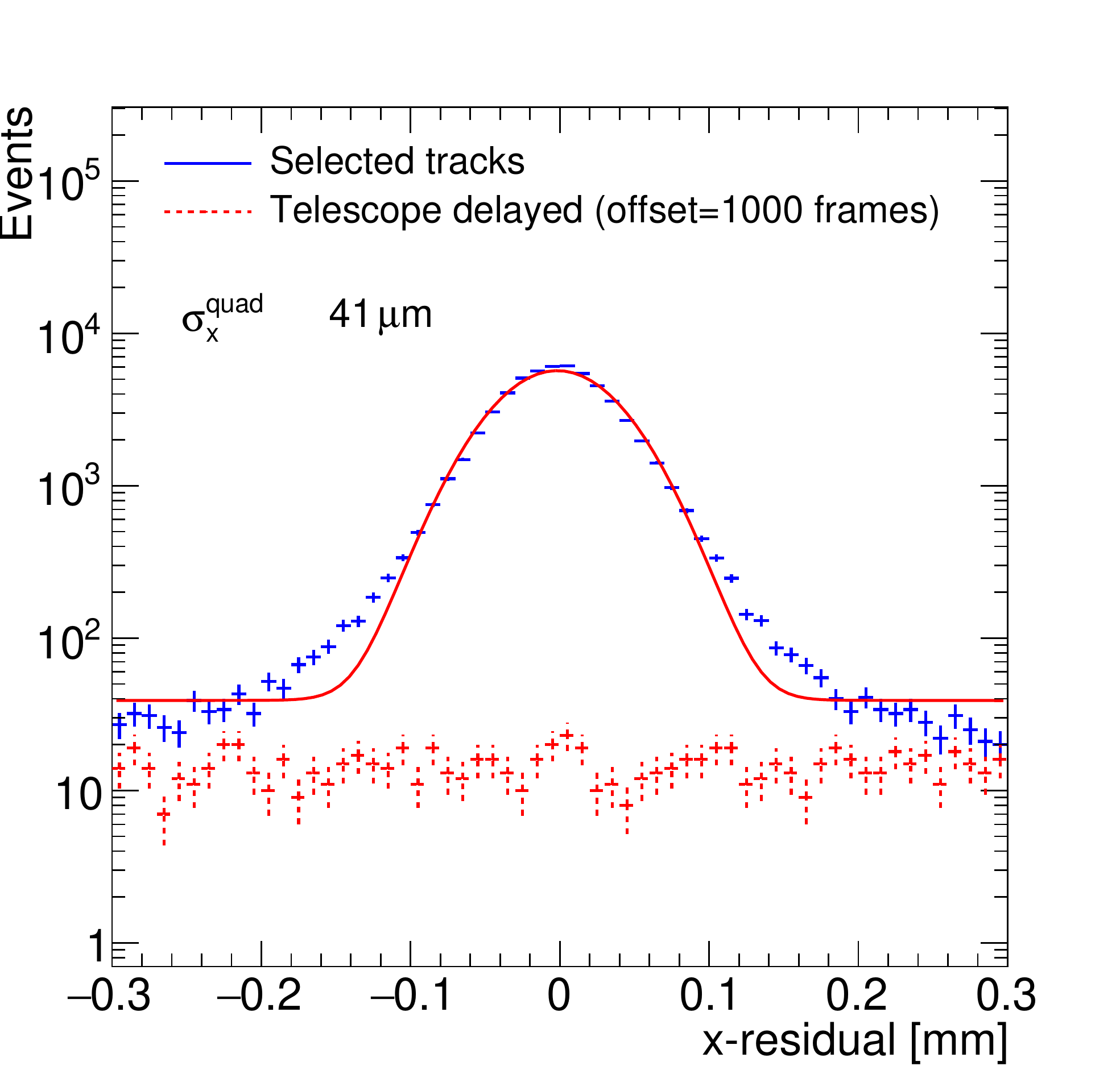}
    \fi
    \caption{difference between mean hit position and track prediction in the pixel plane, fitted with a Gaussian plus a flat background (solid red line). In comparison the background tracks are shown (dashed red), acquired by offsetting the telescope by 1000 frames. }
    \label{fig:outOfSyncBG}
\end{figure}

Figure \ref{fig:outOfSyncBG} displays the difference between the mean position of all quad track hits and the telescope track in the fiducial region. The distribution has long tails, which are in part from unrelated background tracks that are erroneously matched.
The number of background tracks is estimated by shifting the telescope timing by 1000 frames and is shown for comparison. In the fit, these tracks are accounted for by introducing a constant offset.

A fit with a Gaussian function with constant offset yields a standard deviation $\sigma_x^\text{quad}$ of \SI{41}{\um}. This value is the result of various contributions. Firstly, the statistical precision of a position measurement is acquired from a track fit of the quad hits with hit errors. This statistical precision of the position at the center of the quad is \SI{25}{\um}. Furthermore, there is a systematic deviation of \SI{9}{\um} in the pixel plane in the fiducial region after the correction as discussed in section \ref{sec:deformationsPixelPlane}. Additionally, there is a systematic deviation in the $x$-direction of \SI{17}{\um} along the drift direction most likely due to electric field inhomogeneities. This is the $x$-deviation as a function of $z$-position. which should not be confused with the $z$-deviation as a function of $x$ and $y$-position that was mentioned before in section \ref{sec:deformationsDrift}.

In addition, the precision is limited due to multiple scattering in the setup. The precision was calculated with a simple simulation of the setup using the approach to multiple scattering suggested by reference \cite{Patrignani:2016xqp}. The setup has multiple scattering contributions from the telescope planes (0.075\% $X_{0}$ per plane) \cite{Jansen2016}, the air (0.084\% $X_{0}$), the TPC gas (0.09\% $X_{0}$) and the two Kapton foils (0.035\% $X_{0}$) \cite{Patrignani:2016xqp}. By comparing the track angle in the first three telescope planes and the second three telescope planes, the total radiation length of the setup is estimated to be 0.82\% $X_{0}$ (0.66\% $X_{0}$ expected). From the simulation the multiple scattering contribution at the position of the quad center is estimated to be \SI{22}{\um}.

An overview of the contributing errors is given in table \ref{tab:errors}. In the end, there is still a small unidentified systematic error of \SI{14}{\um}.

\begin{table}
    \centering
    \caption{Overview of the errors on the difference between mean hit position and track prediction in the pixel plane}
    \vspace{3mm}
    \begin{tabular}{l r}
        \toprule
        Observed standard deviation $\sigma_x^\text{quad}$ & \SI{41}{\um} \\ \midrule
        Statistical quad detector error & \SI{25}{\um} \\
        Statistical telescope error & \SI{2}{\um} \\
        Systematics over the pixel plane (corrected) & \SI{9}{\um} \\
        Systematics along the drift direction & \SI{17}{\um} \\ 
        Multiple scattering contribution & \SI{22}{\um} \\ \midrule
        Remaining systematic error & \SI{14}{\um} \\ \bottomrule
    \end{tabular}
    \label{tab:errors}
\end{table}

\section{Conclusion and outlook}
A quad module with four Timepix3 based GridPixes has been designed and realised. The module has dimensions of \SI{39.6 x 28.38}{mm} and an active surface of 68.9\%.
The quad module was embedded in a TPC detector and operated at the ELSA test beam facility. 
The single electron resolutions in the transverse and longitudinal planes are similar to the results obtained for the single-chip detector \cite{Ligtenberg:2018a} and primarily limited by diffusion. It is shown that a systematic error from the quad detector for the distortions over the pixel plane of \SI{13}{\um} (\SI{9}{\um} in the central region) has been achieved. The demonstrated resolution of the setup is \SI{41}{\um}, of which the statistical error is \SI{25}{\um}, the error caused by multiple scattering in the setup is \SI{22}{\um} and the total systematic error is \SI{24}{\um}.

The next step is to demonstrate a large detection area with the quad module as a building block and confirm the potential of the GridPix technology for large detectors. A new detector with 8 quad modules carrying a total of 32 Timepix3 Gridpix chips is under construction.

\section*{Acknowledgements}
This research was funded by the Netherlands Organisation for Scientific Research  NWO.  The authors want to thank the support of the mechanical and electronics departments at Nikhef and the accelerator group at the ELSA facility in Bonn. Their gratitude is extended to the Bonn SiLab group for providing the beam telescope.

\bibliography{mybibfile}

\end{document}